\documentclass[
  aps,
  pre,
  a4paper,
  english,
  reprint,
  superscriptaddress
]{revtex4-1}
\usepackage[T1]{fontenc}
\usepackage[utf8]{inputenc}
\usepackage[normalem]{ulem}
\usepackage{graphicx}
\usepackage[american]{babel}
\usepackage{amssymb,amsmath}
\usepackage[
  caption=false
]{subfig}
\usepackage{refstyle}
\usepackage[
  citecolor=blue,
  colorlinks,
  linkcolor=blue,
  urlcolor=blue,
]{hyperref}
\usepackage{xcolor}
\usepackage[babel=true]{csquotes}

\begin{document}

\global\long\def\pgr{\mathcal{P}_{\text{gr}}}
\global\long\def\pdb{\mathcal{P}_{\text{db}}}
\global\long\def\pov{\mathcal{P}_{\text{ov}}}
\global\long\def\pn{\mathcal{P}_{0}}
\global\long\def\df{d_{\text{f}}}

%\title{Hysteretic percolation from locally optimal individual decisions}
%\title{The winner takes it all - How to win globalization?}
%\title{How to win globalization? - Predict economic network formation using node properties}
%\title{From internal properties to topological properties - searching for predictors to the globalization winner}
%\title{Centralities and monopoly - How globalization winner be predicted?}
%\title{How to win globalization? - Predicting global suppliers from network centralities}
%\title{Predicting globalization from network centralities - Who wins and how?}
%\title{}

\title{The winner takes it all - How to win network globalization}

\author{Chengyuan Han}
\email{ch.han@fz-juelich.de}
\affiliation{Forschungszentrum J\"ulich, Institute for Energy and Climate Research - Systems Analysis and Technology Evaluation (IEK-STE), 52428 J\"ulich, Germany }
\affiliation{Institute for Theoretical Physics, University of Cologne, 50937 K\"oln, Germany }

\author{Dirk Witthaut}
\email{d.witthaut@fz-juelich.de}
\affiliation{Forschungszentrum J\"ulich, Institute for Energy and Climate Research - Systems Analysis and Technology Evaluation (IEK-STE), 52428 J\"ulich, Germany }
\affiliation{Institute for Theoretical Physics, University of Cologne, 50937 K\"oln, Germany }

%\author{Jan Nagler}
%\email{jnagler@ethz.ch}
%\affiliation{
%   ETH Z\"urich, Wolfgang-Pauli-Strasse 27, CH-8093 Z\"urich, Switzerland
%}
%\affiliation{
%Computational Social Science, Department of Humanities, Social and Political Sciences,
%ETH Zurich, Clausiusstrasse 50, CH-8092 Zurich, Switzerland
%}
%

\author{Marc Timme}
\email{marc.timme@tu-dresden.de}
\affiliation{
  Chair for Network Dynamics,
  Center for Advancing Electronics Dresden (cfaed) and Institute for Theoretical Physics, TU Dresden
  01062 Dresden, Germany
}

\author{Malte Schr\"oder}
\email{malte.schroeder@tu-dresden.de}
\affiliation{
  Chair for Network Dynamics,
  Center for Advancing Electronics Dresden (cfaed) and Institute for Theoretical Physics, TU Dresden
  01062 Dresden, Germany
}

\begin{abstract}
\noindent Quantifying the importance and power of individual nodes depending on their position in socio-economic networks constitutes a problem across a variety of applications. Examples include the reach of individuals in (online) social networks, the importance of individual banks or loans in financial networks, the relevance of individual companies in supply networks, and the role of traffic hubs in transport networks. 
Which features characterize the importance of a node in a trade network during the emergence of a globalized, connected market? Here we analyze a model that maps the evolution of trade networks to a percolation problem. In particular, we focus on the influence of topological features of the node within the trade network. Our results reveal that an advantageous position with respect to different length scales determines the success of a node at different stages of globalization and depending on the speed of globalization.
%We study the importance of nodes in 
%%%We analyze 
%%%a recently introduced 
%a model of economic percolation that describes 
%%%a game-theoretic network supply problem 
%the game-theoretic evolution of trade networks from a basic network supply problem
%based on individual optimization of the nodes in the network. 
%%%We study the transition from individual, local suppliers to a single, global supplier. 
%In particular, we focus on the question which topological and model parameters determine the importance of a node and how the importance of a node changes depending on these parameters.
%%%in this model, that means the chance of a node to become the final supplier. 
%Our results reveal that different length scales are important for different parameter values and at different stages of the model.
\end{abstract}

\maketitle

\section{Introduction}

Global connectivity
%Connectivity is %a central driving force of 
is central to our
social, economic and technical development \cite{newman03_network_review, albert02_network_review, easley10_networks_crowds_markets, havlin12_challenges}. The growth of a global transportation network has dramatically changed world economy and led to increased efficiency and more centralized production \cite{krugman91_geography}. But %the increasing 
this global connectivity also bears new, systemic risks - highlighted in particular in the financial sector \cite{schweitzer09_economic, piccardi18_complexity}.

Economies of scale are a major driving force in the formation of many of these socio-economic networks. Generally, a well developed economic agent with high connectivity is more attractive or competitive compared to smaller, less developed agents. The larger agents thus naturally attract even more connections \cite{kumar00_scalefreeCopymodel, kumar10_onlineSocial, molkenthin18_socialAdhesion}. In social network theory, this principle is commonly referred to as preferential attachment, driving the formation of scale-free networks \cite{barabasi99_scalefree}. In economic theory, economies of scale have been identified as a key mechanism leading to the emergence of trade networks and globalization \cite{krugman91_geography, krugman91b_geography}. More recently, we have seen the emergence of quasi-monopolies in digital platform economies where economies of scale are particularly strong \cite{katz1994systems, shapiro1998information, brousseau2007economics}. In this case the winner takes it all. But who wins and how? 

%Numerous studies have analyzed the emergence of social or trade networks, but important open questions remain. 
Understanding which node in a network is the most important one and how it `wins' over the competition in a network globalization process is still largely an open question. In particular, a systematic study of network formation in a heterogeneous geographic environment is a demanding task. 
Percolation models describing network growth typically involve random processes \cite{stauffer_92_percolation_book, saberi15_percolationReview, dsouza15_review}, while optimization models of the network structure typically start from a single global objective function \cite{sole01_ecological, milo2002network, gastner06_distribution, memmesheimer06_designingNeural, katifori16_optimal}. 
%Both model classes do not fully apply to 
However, neither model class fully describes 
socio-economic networks, whose formation is determined by the individual decisions (optimization, non-random) of interacting agents (multiple different objective functions). Economic equilibrium models and game-theoretic models capture these interactions and the individual decision but quickly become intractable as the number of agents increases \cite{bala00_noncooperative, jackson02_network_evolution, even07_smallWorldGameTheory, jackson08_network_book, easley10_networks_crowds_markets, atabati15_strategic}. 

In this article, we study a supply network model that explicitly includes nonlinear nonconvex economies of scale and transportation costs and simultaneously enables a semi-analytical treatment by mapping the evolution of the trade network to a percolation problem \cite{schroeder18_individualDecisions}. In the model, agents try to satisfy a given demand at minimum costs, either through domestic production or via imports. Economies of scale favor the centralization of production and the emergence of trade. On the other hand, non-zero transportation costs favor distributed production. %The model is effectively described by a bond percolation problem which allows for an efficient numerical solution even for very large systems.
%These simulations 
Simulating the evolution of the trade network in this model allows us to systematically study how globalization takes place, how the transportation network affects globalization, and last but not least which geographic factors provide an advantage for the economic agents. In particular, we demonstrate that the way to `win' the globalization process is to be in an advantageous position on the correct length scale. We show that the length scale characterizing the competitiveness of a node changes depending on the stage of globalization and the speed of globalization process.

\section{Methods}

\subsection{Economic percolation model} 

\begin{figure}[tb]
\centering
\includegraphics[width=\columnwidth]{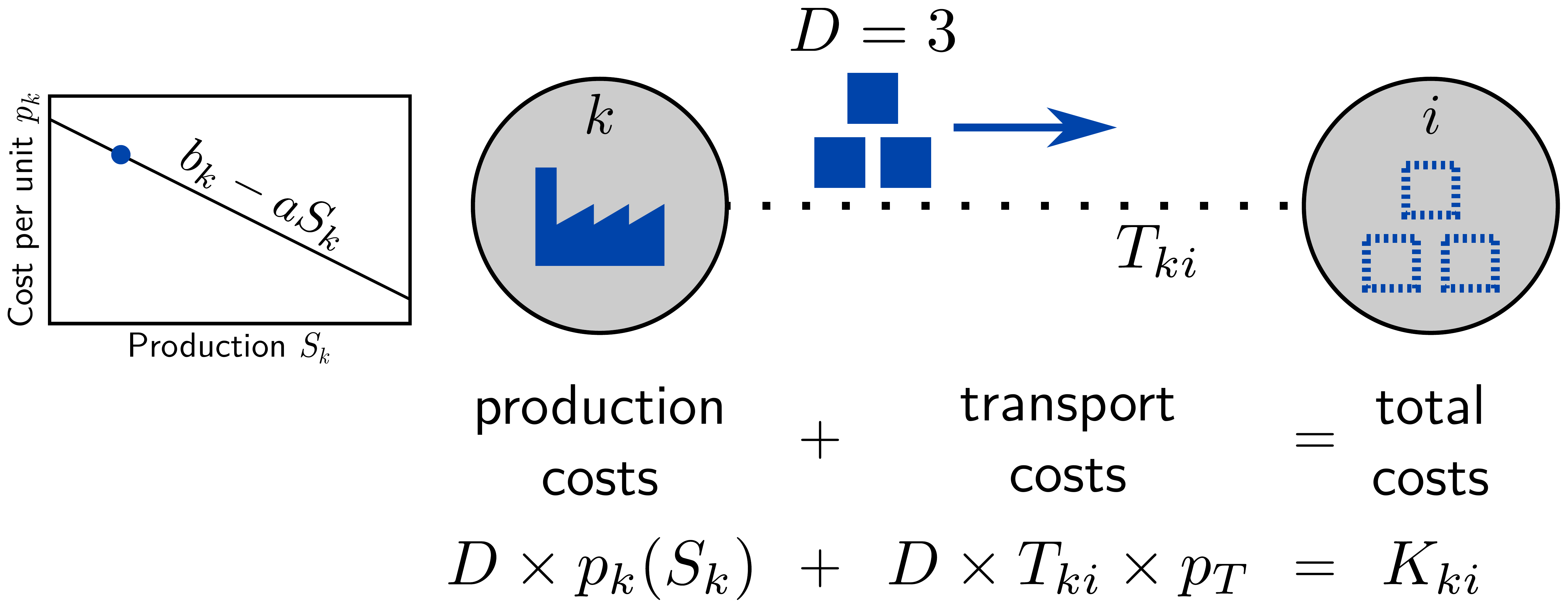}
\caption{
\textbf{Network Supply Problem.} 
Each node $i$ chooses a supplier $k$ to satisfy its demand $D$ at minimal cost $K_{i} = \mathrm{min}_k \, K_{ki}$. These costs include: (I) production costs at node $k$, where the costs per unit depend on the total amount of production $S_k$ at that node (left panel), and (II) transport costs that depend on the distance $T_{ki}$ between the nodes $k$ and $i$ in the underlying transport network (dashed line). All nodes in the network (including $k$) simultaneously solve their individual optimization problem.
}
\label{fig:model_sketch}
\end{figure}

We analyze the influence of topological features on the importance of nodes in a network formation model recently introduced by Schr\"oder et al.~\cite{schroeder18_individualDecisions}. The model describes the formation of trade interactions based on a fundamental network supply problem \cite{krugman91_geography, krugman91b_geography}. The basic idea is as follows: Each node (or economic agent) $i \in \left\{1,2,\dots,N\right\}$ in the network has a fixed demand $D$ (identical for all nodes). A node $i$ can either fill this demand by domestic production or by making purchases from other nodes it is connected to via the underlying transport network. Filling this demand always incurs costs for node $i$: (I) production costs $K^\mathrm{P}_{ki}$ for production at node $k$, even for domestic production where $k=i$, and (II) transport costs $K^\mathrm{T}_{ki}$ for transport from node $k$ to node $i$ if node $i$ makes purchases from other nodes ($k \neq i$). This general setup is illustrated in Fig~\ref{fig:model_sketch}.

The production costs of goods manufactured at node $k$ and consumed at node $i$ are given by
\begin{equation}
     K^\mathrm{P}_{ki} = p_k(S_k) \times S_{ki} ,
\end{equation}
where $S_{ki}$ denotes the amount of goods produced at node $k$ and consumed at node $i$. The costs per unit $p_k$ are \emph{decreasing} with the total %amount of produced goods 
production $S_k = \sum_{i=1}^{N} S_{ki}$ due to \emph{economies of scale} at node $k$. This means production becomes more efficient for larger quantities. Throughout this article we assume a linear relation 
\begin{equation}
   p_k(S_k) = b_k - a S_k \label{eq:scale_efects}
\end{equation}   
for the sake of simplicity, where the parameter $a \ge 0$ directly quantifies the strength of the economies of scale and $b_k$ is a constant offset different for each node, describing inherent production cost advantages. %We assume that the production $S_k$ of a node is given as the sum over all purchases made from that node, $S_k = \sum_i S_{ki}$. 

The transport costs 
\begin{equation}
   K^\mathrm{T}_{ki} = p^\mathrm{T} T_{ki} S_{ki}
\end{equation}
are proportional to the amount of purchased goods $S_{ki}$ and the distance $T_{ki}$ between the nodes in the underlying transport network. The proportionality factor $p^\mathrm{T}$ controls the importance of transport costs relative to production costs. In real-world settings, it typically decreases over time due to technological advancements in the transport sector and serves as the main control parameter for the network formation model. Together, the total costs for node $i$ read
\begin{equation}
	K_{i} = \sum_{k=1}^N K_{ki} = \sum_{k=1}^N K^\mathrm{P}_{ki} + K^\mathrm{T}_{ki}
\end{equation}
as illustrated in Fig~\ref{fig:model_sketch}. 

\begin{figure*}[tb]
\centering
\includegraphics[width=0.95\textwidth]{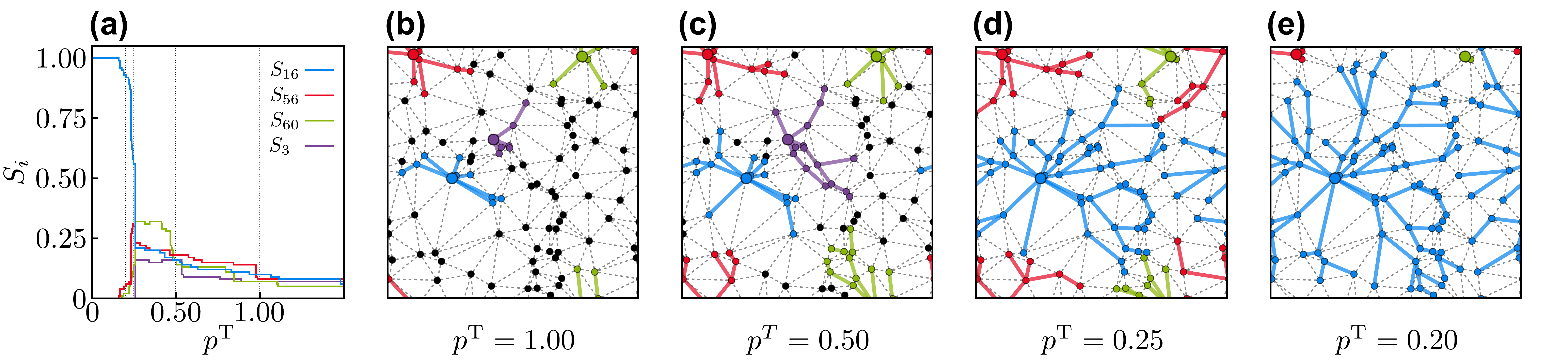}
\caption{
\textbf{Market growth in the percolation model.} (a) Evolution of the size $S_i$ of four markets measured by the production $S_i$ of the markets supplier $i$. Every node in the network optimizes its costs to satisfy its demand as described in the main text. As the importance of transport costs $p^\mathrm{T}$ decreases, nodes make external purchases and common markets (clusters) emerge where production is centralized at a single node $k$. As $p^\mathrm{T} \rightarrow 0$, only a single, global market with supplier $k^* = 16$ and $S_{16} = 1$ remains (blue line). (b-e) Snapshots of the network for different values of $p^\mathrm{T}$. The four markets (clusters) with centralized production shown in panel (a) are illustrated in their respective colors and the central supplier node is highlighted by a larger circle. Black nodes do not belong to any of these four markets. Solid colored lines indicate active links in the transport network, dashed lines indicate potential transport links that are not used by the four large markets.
Parameters are $D = 1/N$, $b \in \left[0,1\right]$ distributed uniformly at random and $a = 10^{-3}$. The planar network is created as the Delaunay triangulation from $N=100$ points distributed uniformly at random in the unit square (see Methods for more details).
}
\label{fig:model_evolution}
\end{figure*}

Each node $i$ chooses its purchases $S_{ki}$ in order to minimize its costs under the constraint that it exactly satisfies its demand, $\sum_k S_{ki} = D$. In general, this leads to $N$ interacting nonlinear and nonconvex optimization problems as the production costs depend on the purchases of all nodes. Nevertheless, a resulting Nash equilibrium, where no node can further decrease its costs by changing its supplier, can be computed efficiently as shown in \cite{schroeder18_individualDecisions}: Each node $i$ chooses only a single supplier $k$ (either itself or one other node in the network) that can be found efficiently with an adapted breadth-first-search. We study the evolution of trade networks starting from the limit of infinite transport costs, $p^\mathrm{T} = \infty$, such that all nodes purchase locally and no trade takes place. As the importance of transport costs decreases, some nodes start to make non-local purchases such that the production $S_k$ of other nodes increases. Eventually, large common markets (clusters) emerge in %the trade network $S_{ki}$,
the network of trades $S_{ki}$,
each with a single supplier node $k$. In the end, when transport costs disappear, $p^\mathrm{T} = 0$, only one giant market (cluster) remains with a single supplier $k^*$ with globally centralized production $S_{k^*} = ND$. This evolution is illustrated in Fig~\ref{fig:model_evolution} for a small planar network. 

In this article we study two main aspects of the formation of this trade network: 
First, how does centralization occur? That is, how does the transition from local production at large $p^\mathrm{T}$ to centralized production at low $p^\mathrm{T}$ take place? Second, we analyze who `wins' the competition. That is, which node $k^*$ becomes the final supplier as production is fully centralized for $p^\mathrm{T} \rightarrow 0$.

\subsection{Analysis of network structure}

The economic percolation model includes heterogeneous geographical conditions explicitly. The matrix $T_{ki}$ encodes the distances of all pairs of nodes $\left(k,i\right)$ which depends on their geographic location and the structure of the underlying transportation network. Hence, the model allows to systematically study the influence of geographical or topological properties on the formation of trade and the centralization of production.
Are there any geographical or topological features that determine which node becomes the final supplier and which does not?

To study the impact of the transport network topology, we consider four different random network ensembles. We start from an ensemble of geographically embedded networks obtained by distributing $N=1000$ nodes uniformly at random on the unit square. Edges are constructed by a Delaunay triangulation with periodic boundary conditions. Each of the resulting $M=3000$ links is undirected and assigned a distance equal to the Euclidean distance between the connected nodes. The distance $T_{ij}$ of two arbitrary nodes $i,j$ in the network is finally obtained as the geodesic or shortest path distance in the network.

The other random network ensembles are obtained from the initial ensemble by a reshuffling of the edges. This procedure keeps the number of connections and the distribution of the individual edge lengths identical and thus leaves the networks comparable to each other. %Three different reshuffling procedures are applied: 
We apply three different reshuffling procedures creating randomizations with different properties: 
First, we keep the structure of the network the same but choose a random permutation of the distances (random weights). This breaks correlations between the link distances and the node position. Second, we uniformly randomly rewire all links to different nodes under the constraint that the resulting network is connected. %(random connections). 
The network then has a topology corresponding to a Poisson random network \cite{albert02_network_review}. Comparison of this randomization to the original network allows us to understand the impact of regular versus random network topologies. 
Third, we create a Barabasi-Albert scale-free network with the same number of links and the same distances for the links \cite{barabasi99_scalefree}. %(random scale-free). 
We thus create four different ensembles with identical average degree and edge lengths, but vastly different global structures. For instance, the degree distribution changes from narrow for the geometric and Poisson random networks to heavy-tailed for for scale-free networks.

\subsection{Model parameters}

In addition to the structure of the transportation network, several model parameters determine the course of globalization. First, we note that the system evolution is invariant with respect to a rescaling of the costs. In particular, we can set $D=1/N$ by choosing an appropriate unit system. A rescaling of the distances can be absorbed into the main control parameter $p^\mathrm{T}$ describing the transport cost per unit. It characterizes the \emph{relative} importance of transportation costs with respect to production costs. 

Two parameters $a$ and $b$ characterize the production costs via the costs per unit $p(S_k) = b_k - a S_k$ [Eq.~(\ref{eq:scale_efects})]. Since only the relative ordering of the costs are relevant to compare different suppliers (in the form of $K_{ki} < K_{ji}$), we scale the costs such that all $b_i \in \left[0,1\right]$ with $\mathrm{min}_i \, b_i = 0$ and $\mathrm{max}_i \, b_i = 1$. In particular, we choose the $b_i$ uniformly at random from the interval $[0,1]$. The second parameter $a$ characterizes the economies of scale and has a strong impact on the model behavior. We perform simulations for vastly different values $a \in \left\{10^{-5},10^{-4}, \dots 10^{1}\right\}$ to cover different regimes of globalization. To put this into context, note that total centralization of production leads to a decrease of production costs by exactly $NDa = a$ for $D = 1/N$. Economies of scale are negligible if $a$ is much smaller than typical differences of the cost parameter $b_i$, i.e., for $a \ll 1/N = 10^{-3}$. Economies of scale are dominant if $a$ is of the order of the largest difference of the $b_i$, i.e. for $a \approx 1$. The range $a \in \left\{10^{-5},10^{-4}, \dots 10^{1}\right\}$ covers both regimes.

In summary, we perform simulations for four different transportation network ensembles and several values of $a$. For each case we consider $1000$ different random realizations of the network with $10$ different permutations of the $b_i$ each, resulting in $10.000$ measurements per ensemble and value of $a$. For each realization, we start the simulation in the limit of large transport costs, $p^\mathrm{T} = \infty$, without any trade interactions. We then gradually lower $p^\mathrm{T}$ and record the emergence of a trade network, i.e., the emergence of connected components of the network defined by the purchases $S_{ki}$, as well as the final supplier for $p^\mathrm{T}=0$. 

\section{Results}

\subsection{How does globalization emerge?}

To understand the emergence of a globalized market we record the size of the largest markets as the transport costs decrease from $p^\mathrm{T} = \infty$ (no trade) to $p^\mathrm{T} = 0$ (single, global market). A trade network between nodes emerges as transportation costs decrease. An example of the globalization of production is shown in Fig \ref{fig:model_evolution} for a small geographically embedded random network. For $p^\mathrm{T} = 1.0$, several nodes have already decided to purchase their goods from other neighboring nodes and multiple markets have formed where production is centralized to a single node. The markets grow when $p^\mathrm{T}$ decreases to $p^\mathrm{T} = 0.5$ as further nodes decide to purchase non-locally. Finally, many nodes again change their supplier, joining one large, global market with strong economies of scale instead of the smaller local markets. In the end, as $p^\mathrm{T} \rightarrow 0$, production is fully centralized at a single node. The size of the four largest markets is shown in Fig \ref{fig:model_evolution} (a) as a function of the transportation cost parameter $p^\mathrm{T}$.

\begin{figure*}[t]
\centering
\includegraphics[width=0.8\textwidth]{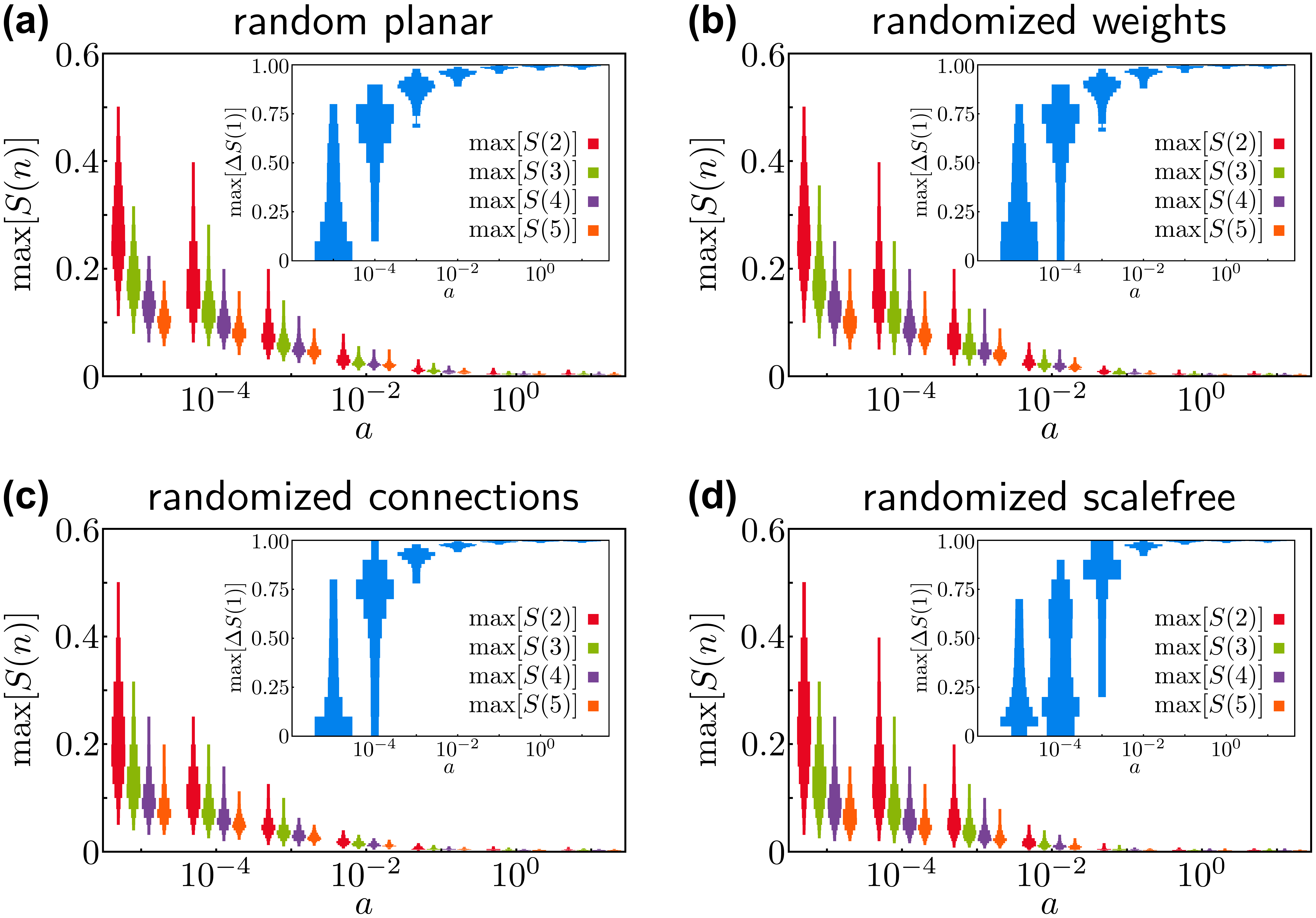}
\caption{
\textbf{Multiple markets or sudden growth?} Distribution of the maximum size $\mathrm{max}[S(n)]$ of the $n$-th largest market and largest change $\mathrm{max}[\Delta S(1)]$ in the size of the largest market (insets) during the globalization process for (a) the random planar network, (b) the network with randomized weights, (c) the network with uniformly randomized links and (d) the network with scale-free randomized links. For small $a$, multiple large clusters appear and merge slowly in all networks. For large $a$, a global market suddenly forms from the individual nodes in a single large cascade before any other market had the chance to grow significantly. Depending on the value of the parameter $a$, nodes have to be competitive at different length scales to become the final supplier. The maximal size of the second largest cluster $\mathrm{max}[S(2)]$(red) can serve as a proxy for this length scale.
}
\label{fig:largest_clusters}
\end{figure*}

Inspecting this evolution, we are directly led to the question how the transition to globalization takes place under different circumstances. Is it very sudden with a single large change in the size of the largest market or is the transition slow and the largest market grows gradually as $p^\mathrm{T}$ decreases? Does a single node expand its market or do multiple large markets grow and only later merge to one global market? To answer these questions, we measure the largest gap $\mathrm{max}[\Delta S(1)]$ in the size (total production) of the largest market \cite{nagler11_single_links} as well as the maximum size of the second largest market $\mathrm{max}[S(2)]$, the third largest market $\mathrm{max}[S(3)]$ and so on over the course of the evolution from infinite to zero transport costs (see Fig~\ref{fig:largest_clusters}). The maximal size $\mathrm{max}[S(2)]$ of the second largest market in particular measures how much markets grow before global centralization occurs. If it is small, only a single large market emerges and local competitiveness is relevant to gain an early advantage. If it is large, at least two large markets expand side by side before one of them becomes globally dominant and production is completely centralized. Here, the markets have to compete against each other on a larger length scale. The maximal size $\mathrm{max}[S(2)]$ of the second largest market serves a proxy for this length scale.

If economies of scale are weak (small values of $a$), multiple large markets coexist before they finally merge. As $a$ becomes larger, the maximum size of all markets except the largest one decreases. Finally, for strong economies of scale $a$, only a single market grows. Correspondingly, globalization becomes more and more abrupt with increasing $a$, measured by the growth of the gap $\mathrm{max}[\Delta S(1)]$. We thus obtain the following picture: For weak economies of scale, several markets grow and finally merge in a 
%continuous 
gradual process. For strong economies of scale, only local markets exist until globalization sets in abruptly. After this sudden transition, exactly one large market remains.

We observe rather little differences between the four network ensembles under consideration. The transition from gradual to abrupt globalization is qualitatively the same in all networks and also the transition point is remarkably similar. While the transition is gradual (no large gaps) for $a=10^{-5}$, it is sudden for $a=10^{-3}$ for all networks. Slight differences are observed only for $a=10^{-4}$. While the maximum gap is larger than $0.1$ for all realization of the random planar network, the transition is still gradual with smaller changes of the largest cluster for most realizations of a scale-free network.

This is rather surprising, as scale free networks are characterized by the existence of hubs, a few nodes with very high degree. At first glance, one might expect that these hubs can exploit economies of scale most easily, making the transition abrupt already for small $a$. Our results show that this simple reasoning fails. The impact of economies of scale on the transition and on the competitiveness of nodes is more subtle. In fact, different hubs have to compete in the globalization process when the economies of scale are not dominant (small $a$). Thus, while hubs allow for the easier formation of local markets, these markets then have to compete on a larger length scale (measured by the maximum size of the second largest market), where the local properties of the central supplier, such as the high degree of the hubs, are less important. Overall, this competition slows down the centralization of production in scale-free networks.

\subsection{Who wins globalization?}

Understanding \emph{how} globalization occurs, we now address the question \emph{who} wins the competition in the current model. % of globalization. 
That is, which node $i$ becomes the final supplier of the network for $p^\mathrm{T} \rightarrow 0$? Are there any geographic features which determine a node's competitiveness? 

%Analyzing \emph{how} to win globalization, we already discussed the impact of hubs in scale-free networks. We now ask the question if it is possible to identify general topological features that determine a node's success to win the competition.

To characterize the geographical location of a node in a network, we consider several different centrality measures that measure different aspects of a node's position in the network:
\begin{itemize}
	\item[(i)]  cost centrality $1/b_i$
	\item[(ii)] local closeness centrality $1 / \mathrm{min}_j T_{ij}$
	\item[(iii)] global closeness centrality $1/ \sum_j T_{ij}$	\cite{sabidussi66_closeness, newman_2010_networks}
	\item[(iv)] degree centrality \cite{newman_2010_networks}
	\item[(v)]  betweenness centrality \cite{freeman77_betweeness, newman_2010_networks}.
\end{itemize}
These quantities measure the advantage of the nodes in terms of (i) global production costs, (ii) small transport costs to a local trade partner, (iii) small transport costs to the whole network, (iv) immediate access to different trade partners and (v) position of the node along many trade routes.

We generally expect that all these properties are beneficial for the nodes. For example, a high cost centrality implies that production is cheap -- at least until production costs decrease significantly due to economies of scale. The node with the highest cost centrality would be the socially optimal supplier when $p^\mathrm{T} = 0$ and minimize the total costs across all nodes. Similarly, a high global closeness centrality implies that transportation is cheap on average, making the node an attractive global supplier when transport costs are not zero. The remaining three centrality measures also point to a favorable position in the network, but their implication is less clear. High degree and local closeness point to an attractive local environment, while high betweenness centrality is a typical measure of importance in social networks and means that many shortest transportation routes cross the respective node.

\begin{figure*}
    \centering
    \includegraphics[width = 1.0\textwidth]{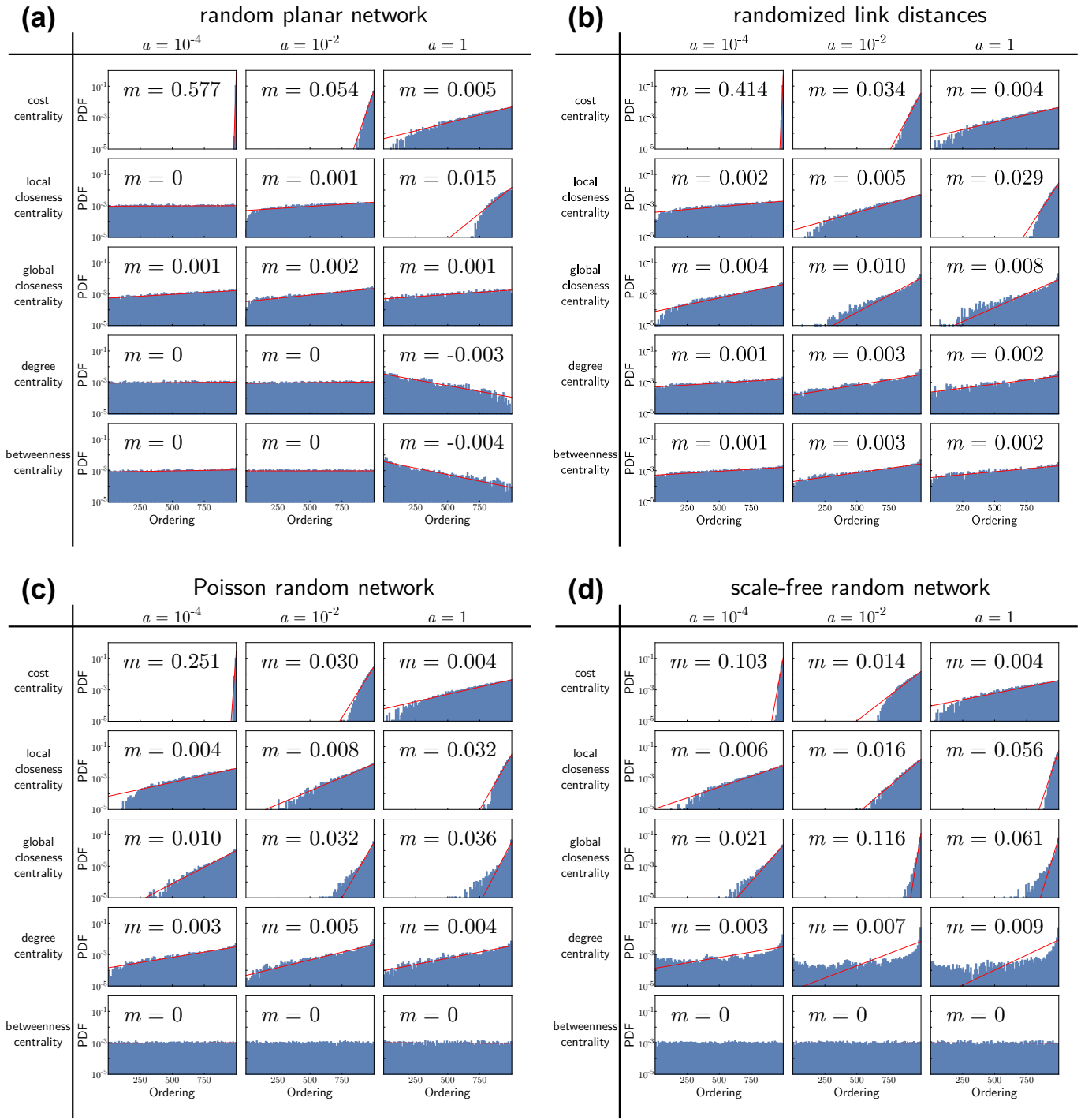}
    \caption{\textbf{How to win network globalization?} Distribution of the ranking of the final supplier in various centrality measures (see main text) in (a) a random planar network, (b) the network with a random permutation of edge distances, (c) a Poisson random network with a random permutation of the edge distances, and (d) a scale-free network with a random permutation of the edge distances. All networks are constructed from a Delaunay triangulation of $N=1000$ points uniformly randomly distributed in the unit square, resulting in $M=3000$ links with distances equal to the Euclidean distance between the connected nodes (see Methods for details).}
    \label{fig:histogram}
\end{figure*}

To understand which of these properties %is truly important for competitiveness, 
most strongly influences the competitiveness of a node, 
we rank all nodes according to their centralities and evaluate if the final suppliers typically have a high or low ranking. We record the final supplier and its centrality ranking $x$ for each random realization of the globalization process. The resulting distributions of the ranks of the final supplier are shown %in figures \ref{fig:histogram_delaunay},\ref{fig:histogram_delaunay_randomized_weights},\ref{fig:histogram_delaunay_randomized_connections} and \ref{fig:histogram_delaunay_randomized_scalefree}
in Fig~\ref{fig:histogram} 
for the four network ensembles under consideration. In addition, we fit a distribution $P(x) \sim \exp\left[ -m (N - x) \right]$ to the observed centrality rankings to quantify the importance of the respective centrality. A value of  $m = 0$ indicates a flat distribution, i.e., no influence of the centrality rank $x$ on the chance to become the final supplier. The higher the value of $|m|$, the stronger the correlation, and the more meaningful the respective centrality to predict which node wins the competition. 

The first, expected observation is the influence of the cost centrality $1/b_i$ of a node $i$. For weak economies of scale (small $a$) the production costs are dominated by the cost parameters $b_i$ and low production costs are decisive for the competitiveness of a node. For all network ensembles under consideration, cost centrality is the best indicator for competitiveness for small $a$, whereas its importance decreases for stronger economies of scale. 

The second, more striking observation is the importance of the local closeness centrality. In the case of strong economies of scale $a=1$, this centrality measure provides the best indicator for the competitiveness of a node. The histogram of the centrality ranking peaks strongly at top ranks. Local closeness is even more important than global closeness, although we evaluate the global competitiveness of the nodes. Again, this finding holds true for all four network ensembles.

A surprising correlation is found for the two remaining centrality measures, degree and betweenness, for the spatially embedded random network. Contrary to our expectation, the final supplier typically has a \emph{low} degree and betweenness centrality for strong economies of scale $a$. This effect is lost or even reversed for the other network ensembles and can be attributed to a subtle geometric property of spatially embedded random networks. In this network class, local closeness centrality is anti-correlated with degree and betweenness centrality. As competitive nodes have a high local closeness, they are likely to have a low degree and betweenness centrality. This observation is particularly relevant since real-world transportation networks are typically spatially embedded, with the exception of digital, data exchange networks. ote that similar correlations exist for other network ensembles as well. For example, nodes with a high degree centrality in the reshuffled scale free networks typically also have high local closeness centrality, due to more opportunities for a short link.

\begin{figure*}
\centering
\includegraphics[width=0.8\textwidth]{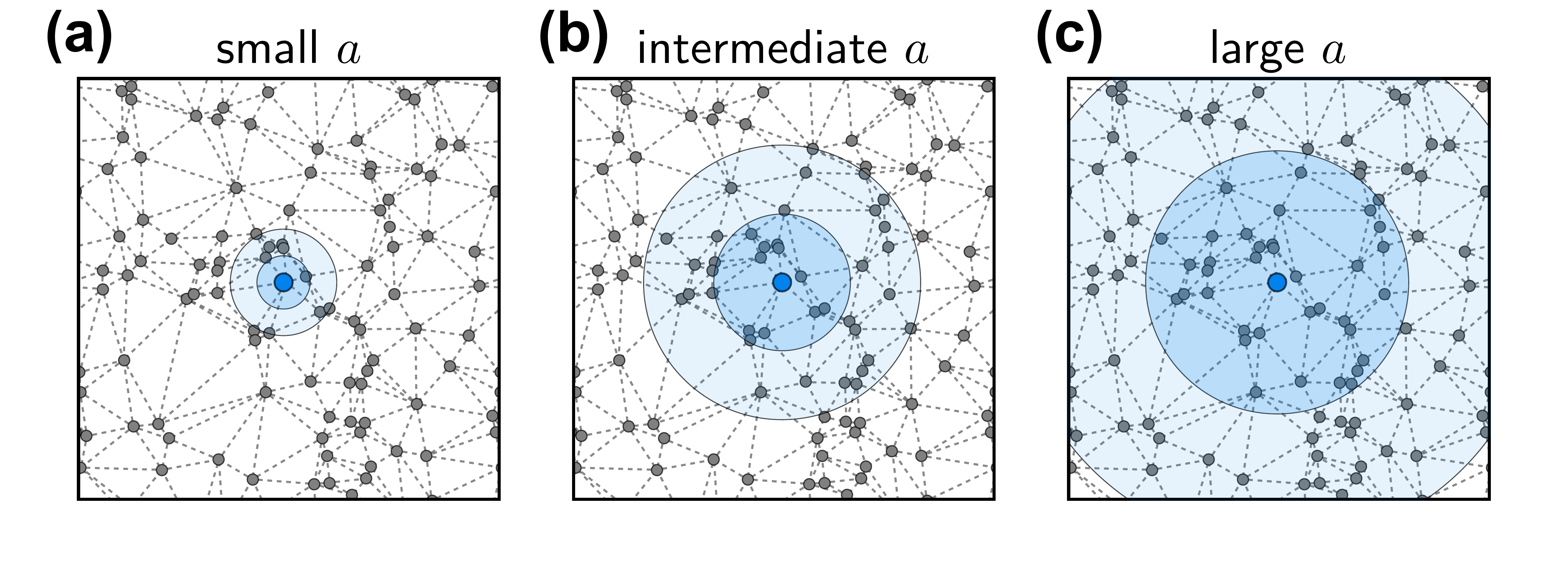}
\caption{
\textbf{Impact of a single customer.} Sketch of the effect of a single (new) customer for a node. With the new customer production increases and the production costs per unit decrease by $a \, D$. This compensates larger transport costs for nodes further away from the supplier. Consequently, the supplier becomes competitive in a larger range and can potentially attract additional customers. The blue circles indicate the distance that is compensated by the decrease in production costs due to one customer (two customers). (a) For small $a$, the change in production cost is small and likely has no immediate effect [compare $a = 10^{-4}$ in Fig~\ref{fig:histogram}(a)]. The nodes have to compete at all length scales. (b) For intermediate $a$, a single customer may reduce the costs sufficiently to cause additional nodes to change their supplier. In this case, nodes have to compete at a local scale until they reach a size sufficiently large to take over the global market. (c) For large $a$, a single customer definitely reduces the costs sufficiently to cause a cascade of purchasing decisions and the first node to attract a customer takes over the whole market. Here, only the immediate neighborhood of a node decides about its success [compare $a = 1$ in Fig~\ref{fig:histogram}(a)].
}
\label{fig:model_schematic}
\end{figure*}

Finally, a more subtle implication of the centrality measures is that, depending on the parameter $a$, the size or length scale of the relevant neighborhood changes. This length scale is defined by the critical size the largest cluster must reach before it becomes the global supplier. The effect is illustrated in Fig~\ref{fig:model_schematic}. For small $a$, the number of customers does not affect the costs very much and one new customer allows the supplier to attract customers only in a small additional range [Fig~\ref{fig:model_schematic} (a)]. Consequently, a node must attract a larger number of customers to become globally competitive and the critical size is (almost) equal to the total size of the network. In this regime, global centrality measures like the cost centrality are most relevant. For intermediate $a$, a single customer allows the supplier to attract nodes in a larger range [Fig~\ref{fig:model_schematic} (b)]. The critical length scale becomes smaller and we need to put more weight to the local structure. In this regime, the global closeness centrality and the degree centrality start to become better predictors, quantifying the centrality of a node in a local neighborhood. Finally, for very large $a$, the critical size of the largest cluster becomes $2$ and one single customer induces a sufficiently large change in production costs for the supplier to become globally competitive [Fig~\ref{fig:model_schematic} (c)]. The centrality of a node in its most local context then becomes the deciding factor. This is best measured by the distance to the nearest neighbor, the local closeness centrality $1 / \mathrm{min}_j T_{ij}$. 

Comparing results across the different network topologies, we find that the network topology becomes more important when the diameter is smaller, i.e., for Poisson and scale-free network structure. Since the total transport costs in these networks are smaller (proportional to the smaller diameter of these networks), the critical size to become the global supplier is also smaller. Thus, local length scales and the (local) network structure become important already for smaller values of $a$. 

%Loosely speaking, we can summarize our findings as follows: In the case of weak economies of scale the internal properties of a node are decisive. The most efficient nodes, characterized by small production costs $b_i$, compete for the global market. For strong economies of scale, speed becomes the most important factor, %more important then 
%rather than 
%efficiency or location. A node with high local closeness can rapidly attract the first external customers and then exploit economies of scale to grow its market. Hence, this node has an advantage that other nodes can hardly compensate. 

\section{Conclusion and Discussion}

Economies of scale are a decisive factor in the formation of socio-economic networks and the globalization and centralization of economic activities. Eventually, the winner takes all. Here we have studied core aspects of the question who wins and how.

The formation of socio-economic networks is a guiding research question across disciplines, including economics \cite{krugman91_geography, krugman91b_geography, schweitzer09_economic, havlin12_challenges}, sociology \cite{jackson08_network_book, easley10_networks_crowds_markets, vanDolder14_individual} and statistical physics \cite{barabasi99_scalefree, albert02_network_review}. Key mechanisms and global properties of network formation through economies of scale have been thoroughly analyzed \cite{krugman91_geography, barabasi99_scalefree, jackson08_network_book}, whereas the microscopic processes in large systems with many heterogeneous actors are much harder to grasp. Most traditional models of network formation do not explicitly capture the behavior of individual actors \cite{bertsekas98_optimization, barabasi99_scalefree, saberi15_percolationReview}. Percolation models are based on random processes, while optimization models typically assume a common global objective function. In contrast, game theoretic models describing individual agents \cite{watts01_dynamic, jackson02_network_evolution, gastner06_distribution, even07_smallWorldGameTheory} are typically hard, if not impossible, to solve for large heterogeneous systems. In this article, we have analyzed a supply network model \cite{schroeder18_individualDecisions} that explicitly includes economies of scale and individual decisions, yet remains simple enough to allow for an efficient simulation of  network formation and centralization in large heterogeneous environments. We exploit this fact %to systematically study the question of who wins the globalization process and how. 
to reveal the most important topological properties that determine the success of a node in the globalization process. We find that different length scales determine the success of a node, depending on the speed of the globalization.

The model yields the structure of a trade network given an underlying transportation network as a function of two main parameters: the strength of economies of scale $a$ and the transport costs per distance $p^\mathrm{T}$. As transport costs decrease, trade links are established and the production is centralized to fewer and fewer nodes. For weak economies of scale, this process is gradual. Nodes compete at all length scales and the internal cost parameters are decisive for the competitiveness of a node. Only nodes with low productions costs $b_i$ have a chance to become the final supplier of the network once production is centralized completely. The geographic location of the nodes in the network, characterized by different centrality measures, plays only a minor role. In contrast, if economies of scale become dominant, this picture changes entirely: Production is centralized in a single transition once transportation costs decrease below a critical value. Only a single node attracts a significant number of customers and wins the competition. Moreover, the transition becomes abrupt and as such hard to foresee. The chance of a node to win is now mostly determined by the location of the node in the network. Interestingly, however, global centrality measures are not the best indicator for competitiveness. Instead, a local measure of the distance to the nearest neighbor, referred to as local closeness, is the best indicator for the success of a node. 

Loosely speaking, we our findings are as follows: For weak economies of scale the internal properties of a node or economic agent are decisive. Competition occurs across all length scales in the network and basic efficiency provides the greatest advantage in all stages of globalization. Only the (globally) most efficient nodes have a chance to take over the market. For strong economies of scale speed becomes the most important factor, rather than efficiency or global location. Competition occurs only locally to gain a first advantage and only the agent with the highest local closeness can rapidly attract the first external customers and then exploit economies of scale to grow its market, skipping over the competition in other stages of globalization.
For the future it would be of eminent interest to study how other factors influencing globalization confirm or modify these findings and whether they can be confirmed in real world settings.

\section{Acknowledgements}

We gratefully acknowledge support from the Helmholtz association (grant no. VH-NG-1025 to DW), the German Ministry for Education and Research (BMBF grants no. 03SF0472A-F and 03EK3055 to DW and MT) and the German Science Foundation (DFG) by a grant towards the Cluster of Excellence \emph{Center for Advancing Electronics Dresden} (cfaed).

\section{Author Contributions}

\noindent Conceptualization: Dirk Witthaut and Malte Schr\"oder\\ 	
Formal Analysis: Chengyuan Han and Malte Schr\"oder\\ 	 	
Funding Acquisition: Dirk Witthaut and Marc Timme \\ 	
Investigation: Chengyuan Han and Malte Schr\"oder\\ 	
Methodology: Chengyuan Han, Dirk Witthaut and Malte Schr\"oder\\ 	
Project Administration: Dirk Witthaut and Malte Schr\"oder\\ 	 	
Resources: Chengyuan Han, Dirk Witthaut, Malte Schr\"oder and Marc Timme\\ 	 	
Software: Chengyuan Han and Malte Schr\"oder \\ 	 	
Supervision: Dirk Witthaut and Malte Schr\"oder\\ 	 	
Validation: Chengyuan Han and Malte Schr\"oder\\ 	
Visualization: Malte Schr\"oder\\ 	
Writing -- Original Draft Preparation: Dirk Witthaut and Malte Schr\"oder\\ 	
Writing -- Review and Editing: Chengyuan Han, Dirk Witthaut, Marc Timme and Malte Schr\"oder\\ 	

\bibliography{manuscript_winnerTakesAll}

\begin{thebibliography}{36}%
\makeatletter
\providecommand \@ifxundefined [1]{%
 \@ifx{#1\undefined}
}%
\providecommand \@ifnum [1]{%
 \ifnum #1\expandafter \@firstoftwo
 \else \expandafter \@secondoftwo
 \fi
}%
\providecommand \@ifx [1]{%
 \ifx #1\expandafter \@firstoftwo
 \else \expandafter \@secondoftwo
 \fi
}%
\providecommand \natexlab [1]{#1}%
\providecommand \enquote  [1]{``#1''}%
\providecommand \bibnamefont  [1]{#1}%
\providecommand \bibfnamefont [1]{#1}%
\providecommand \citenamefont [1]{#1}%
\providecommand \href@noop [0]{\@secondoftwo}%
\providecommand \href [0]{\begingroup \@sanitize@url \@href}%
\providecommand \@href[1]{\@@startlink{#1}\@@href}%
\providecommand \@@href[1]{\endgroup#1\@@endlink}%
\providecommand \@sanitize@url [0]{\catcode `\\12\catcode `\$12\catcode
  `\&12\catcode `\#12\catcode `\^12\catcode `\_12\catcode `\%12\relax}%
\providecommand \@@startlink[1]{}%
\providecommand \@@endlink[0]{}%
\providecommand \url  [0]{\begingroup\@sanitize@url \@url }%
\providecommand \@url [1]{\endgroup\@href {#1}{\urlprefix }}%
\providecommand \urlprefix  [0]{URL }%
\providecommand \Eprint [0]{\href }%
\providecommand \doibase [0]{http://dx.doi.org/}%
\providecommand \selectlanguage [0]{\@gobble}%
\providecommand \bibinfo  [0]{\@secondoftwo}%
\providecommand \bibfield  [0]{\@secondoftwo}%
\providecommand \translation [1]{[#1]}%
\providecommand \BibitemOpen [0]{}%
\providecommand \bibitemStop [0]{}%
\providecommand \bibitemNoStop [0]{.\EOS\space}%
\providecommand \EOS [0]{\spacefactor3000\relax}%
\providecommand \BibitemShut  [1]{\csname bibitem#1\endcsname}%
\let\auto@bib@innerbib\@empty
%</preamble>
\bibitem [{\citenamefont {Newman}(2003)}]{newman03_network_review}%
  \BibitemOpen
  \bibfield  {author} {\bibinfo {author} {\bibfnamefont {M.}~\bibnamefont
  {Newman}},\ }\href@noop {} {\bibfield  {journal} {\bibinfo  {journal} {SIAM
  Review}\ }\textbf {\bibinfo {volume} {45}},\ \bibinfo {pages} {167} (\bibinfo
  {year} {2003})}\BibitemShut {NoStop}%
\bibitem [{\citenamefont {Albert}\ and\ \citenamefont
  {Barab{\'a}si}(2002)}]{albert02_network_review}%
  \BibitemOpen
  \bibfield  {author} {\bibinfo {author} {\bibfnamefont {R.}~\bibnamefont
  {Albert}}\ and\ \bibinfo {author} {\bibfnamefont {A.-L.}\ \bibnamefont
  {Barab{\'a}si}},\ }\href@noop {} {\bibfield  {journal} {\bibinfo  {journal}
  {Rev. Mod. Phys.}\ }\textbf {\bibinfo {volume} {74}},\ \bibinfo {pages} {47}
  (\bibinfo {year} {2002})}\BibitemShut {NoStop}%
\bibitem [{\citenamefont {Easley}\ and\ \citenamefont
  {Kleinberg}(2010)}]{easley10_networks_crowds_markets}%
  \BibitemOpen
  \bibfield  {author} {\bibinfo {author} {\bibfnamefont {D.}~\bibnamefont
  {Easley}}\ and\ \bibinfo {author} {\bibfnamefont {J.}~\bibnamefont
  {Kleinberg}},\ }\href@noop {} {\emph {\bibinfo {title} {Networks, crowds, and
  markets: Reasoning about a highly connected world}}}\ (\bibinfo  {publisher}
  {Cambridge University Press, Cambridge},\ \bibinfo {year} {2010})\BibitemShut
  {NoStop}%
\bibitem [{\citenamefont {Havlin}\ \emph {et~al.}(2012)\citenamefont {Havlin},
  \citenamefont {Kenett}, \citenamefont {Ben-Jacob}, \citenamefont {Bunde},
  \citenamefont {Cohen}, \citenamefont {Hermann}, \citenamefont {Kantelhardt},
  \citenamefont {Kert{\'e}sz}, \citenamefont {Kirkpatrick}, \citenamefont
  {Kurths}, \citenamefont {Portugali},\ and\ \citenamefont
  {Solomon}}]{havlin12_challenges}%
  \BibitemOpen
  \bibfield  {author} {\bibinfo {author} {\bibfnamefont {S.}~\bibnamefont
  {Havlin}}, \bibinfo {author} {\bibfnamefont {D.~Y.}\ \bibnamefont {Kenett}},
  \bibinfo {author} {\bibfnamefont {E.}~\bibnamefont {Ben-Jacob}}, \bibinfo
  {author} {\bibfnamefont {A.}~\bibnamefont {Bunde}}, \bibinfo {author}
  {\bibfnamefont {R.}~\bibnamefont {Cohen}}, \bibinfo {author} {\bibfnamefont
  {H.}~\bibnamefont {Hermann}}, \bibinfo {author} {\bibfnamefont
  {J.}~\bibnamefont {Kantelhardt}}, \bibinfo {author} {\bibfnamefont
  {J.}~\bibnamefont {Kert{\'e}sz}}, \bibinfo {author} {\bibfnamefont
  {S.}~\bibnamefont {Kirkpatrick}}, \bibinfo {author} {\bibfnamefont
  {J.}~\bibnamefont {Kurths}}, \bibinfo {author} {\bibfnamefont
  {J.}~\bibnamefont {Portugali}}, \ and\ \bibinfo {author} {\bibfnamefont
  {S.}~\bibnamefont {Solomon}},\ }\href@noop {} {\bibfield  {journal} {\bibinfo
   {journal} {Eur. Phys. J. Special Topics}\ }\textbf {\bibinfo {volume}
  {214}},\ \bibinfo {pages} {273} (\bibinfo {year} {2012})}\BibitemShut
  {NoStop}%
\bibitem [{\citenamefont {Krugman}(1991{\natexlab{a}})}]{krugman91_geography}%
  \BibitemOpen
  \bibfield  {author} {\bibinfo {author} {\bibfnamefont {P.~R.}\ \bibnamefont
  {Krugman}},\ }\href@noop {} {\emph {\bibinfo {title} {Geography and trade}}}\
  (\bibinfo  {publisher} {MIT press, Cambridge, MA},\ \bibinfo {year}
  {1991})\BibitemShut {NoStop}%
\bibitem [{\citenamefont {Schweitzer}\ \emph {et~al.}(2009)\citenamefont
  {Schweitzer}, \citenamefont {Fagiolo}, \citenamefont {Sornette},
  \citenamefont {Vega-Redondo}, \citenamefont {Vespignani},\ and\ \citenamefont
  {White}}]{schweitzer09_economic}%
  \BibitemOpen
  \bibfield  {author} {\bibinfo {author} {\bibfnamefont {F.}~\bibnamefont
  {Schweitzer}}, \bibinfo {author} {\bibfnamefont {G.}~\bibnamefont {Fagiolo}},
  \bibinfo {author} {\bibfnamefont {D.}~\bibnamefont {Sornette}}, \bibinfo
  {author} {\bibfnamefont {F.}~\bibnamefont {Vega-Redondo}}, \bibinfo {author}
  {\bibfnamefont {A.}~\bibnamefont {Vespignani}}, \ and\ \bibinfo {author}
  {\bibfnamefont {D.~R.}\ \bibnamefont {White}},\ }\href@noop {} {\bibfield
  {journal} {\bibinfo  {journal} {Science}\ }\textbf {\bibinfo {volume}
  {325}},\ \bibinfo {pages} {422} (\bibinfo {year} {2009})}\BibitemShut
  {NoStop}%
\bibitem [{\citenamefont {Piccardi}\ and\ \citenamefont
  {Tajoli}(2018)}]{piccardi18_complexity}%
  \BibitemOpen
  \bibfield  {author} {\bibinfo {author} {\bibfnamefont {C.}~\bibnamefont
  {Piccardi}}\ and\ \bibinfo {author} {\bibfnamefont {L.}~\bibnamefont
  {Tajoli}},\ }\href@noop {} {\bibfield  {journal} {\bibinfo  {journal} {PloS
  one}\ }\textbf {\bibinfo {volume} {13}},\ \bibinfo {pages} {e0208265}
  (\bibinfo {year} {2018})}\BibitemShut {NoStop}%
\bibitem [{\citenamefont {Kumar}\ \emph {et~al.}(2000)\citenamefont {Kumar},
  \citenamefont {Raghavan}, \citenamefont {Rajagopalan}, \citenamefont
  {Sivakumar}, \citenamefont {Tomkins},\ and\ \citenamefont
  {Upfal}}]{kumar00_scalefreeCopymodel}%
  \BibitemOpen
  \bibfield  {author} {\bibinfo {author} {\bibfnamefont {R.}~\bibnamefont
  {Kumar}}, \bibinfo {author} {\bibfnamefont {P.}~\bibnamefont {Raghavan}},
  \bibinfo {author} {\bibfnamefont {S.}~\bibnamefont {Rajagopalan}}, \bibinfo
  {author} {\bibfnamefont {D.}~\bibnamefont {Sivakumar}}, \bibinfo {author}
  {\bibfnamefont {A.}~\bibnamefont {Tomkins}}, \ and\ \bibinfo {author}
  {\bibfnamefont {E.}~\bibnamefont {Upfal}},\ }in\ \href@noop {} {\emph
  {\bibinfo {booktitle} {Proceedings 41st Annual Symposium on Foundations of
  Computer Science}}}\ (\bibinfo {year} {2000})\ pp.\ \bibinfo {pages}
  {57--65}\BibitemShut {NoStop}%
\bibitem [{\citenamefont {Kumar}\ \emph {et~al.}(2010)\citenamefont {Kumar},
  \citenamefont {Novak},\ and\ \citenamefont {Tomkins}}]{kumar10_onlineSocial}%
  \BibitemOpen
  \bibfield  {author} {\bibinfo {author} {\bibfnamefont {R.}~\bibnamefont
  {Kumar}}, \bibinfo {author} {\bibfnamefont {J.}~\bibnamefont {Novak}}, \ and\
  \bibinfo {author} {\bibfnamefont {A.}~\bibnamefont {Tomkins}},\ }\enquote
  {\bibinfo {title} {Structure and evolution of online social networks},}\ in\
  \href@noop {} {\emph {\bibinfo {booktitle} {Link Mining: Models, Algorithms,
  and Applications}}},\ \bibinfo {editor} {edited by\ \bibinfo {editor}
  {\bibfnamefont {P.~S.}\ \bibnamefont {Yu}}, \bibinfo {editor} {\bibfnamefont
  {J.}~\bibnamefont {Han}}, \ and\ \bibinfo {editor} {\bibfnamefont
  {C.}~\bibnamefont {Faloutsos}}}\ (\bibinfo  {publisher} {Springer New York},\
  \bibinfo {address} {New York, NY},\ \bibinfo {year} {2010})\ pp.\ \bibinfo
  {pages} {337--357}\BibitemShut {NoStop}%
\bibitem [{\citenamefont {Molkenthin}\ \emph {et~al.}(2018)\citenamefont
  {Molkenthin}, \citenamefont {Schr\"oder},\ and\ \citenamefont
  {Timme}}]{molkenthin18_socialAdhesion}%
  \BibitemOpen
  \bibfield  {author} {\bibinfo {author} {\bibfnamefont {N.}~\bibnamefont
  {Molkenthin}}, \bibinfo {author} {\bibfnamefont {M.}~\bibnamefont
  {Schr\"oder}}, \ and\ \bibinfo {author} {\bibfnamefont {M.}~\bibnamefont
  {Timme}},\ }\href@noop {} {\bibfield  {journal} {\bibinfo  {journal} {Phys.
  Rev. Lett.}\ }\textbf {\bibinfo {volume} {121}},\ \bibinfo {pages} {138301}
  (\bibinfo {year} {2018})}\BibitemShut {NoStop}%
\bibitem [{\citenamefont {Barab{\'a}si}\ and\ \citenamefont
  {Albert}(1999)}]{barabasi99_scalefree}%
  \BibitemOpen
  \bibfield  {author} {\bibinfo {author} {\bibfnamefont {A.-L.}\ \bibnamefont
  {Barab{\'a}si}}\ and\ \bibinfo {author} {\bibfnamefont {R.}~\bibnamefont
  {Albert}},\ }\href@noop {} {\bibfield  {journal} {\bibinfo  {journal}
  {Science}\ }\textbf {\bibinfo {volume} {286}},\ \bibinfo {pages} {509}
  (\bibinfo {year} {1999})}\BibitemShut {NoStop}%
\bibitem [{\citenamefont {Krugman}(1991{\natexlab{b}})}]{krugman91b_geography}%
  \BibitemOpen
  \bibfield  {author} {\bibinfo {author} {\bibfnamefont {P.~R.}\ \bibnamefont
  {Krugman}},\ }\href@noop {} {\bibfield  {journal} {\bibinfo  {journal} {J.
  Pol. Econ.}\ }\textbf {\bibinfo {volume} {99}},\ \bibinfo {pages} {483}
  (\bibinfo {year} {1991}{\natexlab{b}})}\BibitemShut {NoStop}%
\bibitem [{\citenamefont {Katz}\ and\ \citenamefont
  {Shapiro}(1994)}]{katz1994systems}%
  \BibitemOpen
  \bibfield  {author} {\bibinfo {author} {\bibfnamefont {M.~L.}\ \bibnamefont
  {Katz}}\ and\ \bibinfo {author} {\bibfnamefont {C.}~\bibnamefont {Shapiro}},\
  }\href@noop {} {\bibfield  {journal} {\bibinfo  {journal} {J. Econ.
  Perspect.}\ }\textbf {\bibinfo {volume} {8}},\ \bibinfo {pages} {93}
  (\bibinfo {year} {1994})}\BibitemShut {NoStop}%
\bibitem [{\citenamefont {Shapiro}\ \emph {et~al.}(1998)\citenamefont
  {Shapiro}, \citenamefont {Carl},\ and\ \citenamefont
  {Varian}}]{shapiro1998information}%
  \BibitemOpen
  \bibfield  {author} {\bibinfo {author} {\bibfnamefont {C.}~\bibnamefont
  {Shapiro}}, \bibinfo {author} {\bibfnamefont {S.}~\bibnamefont {Carl}}, \
  and\ \bibinfo {author} {\bibfnamefont {H.~R.}\ \bibnamefont {Varian}},\
  }\href@noop {} {\emph {\bibinfo {title} {Information rules: a strategic guide
  to the network economy}}}\ (\bibinfo  {publisher} {Harvard Business Press,
  Brighton, MA},\ \bibinfo {year} {1998})\BibitemShut {NoStop}%
\bibitem [{\citenamefont {Brousseau}\ and\ \citenamefont
  {Penard}(2007)}]{brousseau2007economics}%
  \BibitemOpen
  \bibfield  {author} {\bibinfo {author} {\bibfnamefont {E.}~\bibnamefont
  {Brousseau}}\ and\ \bibinfo {author} {\bibfnamefont {T.}~\bibnamefont
  {Penard}},\ }\href@noop {} {\bibfield  {journal} {\bibinfo  {journal} {Rev.
  Netw. Econ.}\ }\textbf {\bibinfo {volume} {6}} (\bibinfo {year}
  {2007})}\BibitemShut {NoStop}%
\bibitem [{\citenamefont {Stauffer}\ and\ \citenamefont
  {Aharony}(1992)}]{stauffer_92_percolation_book}%
  \BibitemOpen
  \bibfield  {author} {\bibinfo {author} {\bibfnamefont {D.}~\bibnamefont
  {Stauffer}}\ and\ \bibinfo {author} {\bibfnamefont {A.}~\bibnamefont
  {Aharony}},\ }\href@noop {} {\emph {\bibinfo {title} {Introduction to
  Percolation Theory}}}\ (\bibinfo  {publisher} {Taylor \& Francis, London},\
  \bibinfo {year} {1992})\BibitemShut {NoStop}%
\bibitem [{\citenamefont {Saberi}(2015)}]{saberi15_percolationReview}%
  \BibitemOpen
  \bibfield  {author} {\bibinfo {author} {\bibfnamefont {A.~A.}\ \bibnamefont
  {Saberi}},\ }\href@noop {} {\bibfield  {journal} {\bibinfo  {journal} {Phys.
  Rep.}\ }\textbf {\bibinfo {volume} {578}},\ \bibinfo {pages} {1 } (\bibinfo
  {year} {2015})}\BibitemShut {NoStop}%
\bibitem [{\citenamefont {D'Souza}\ and\ \citenamefont
  {Nagler}(2015)}]{dsouza15_review}%
  \BibitemOpen
  \bibfield  {author} {\bibinfo {author} {\bibfnamefont {R.~M.}\ \bibnamefont
  {D'Souza}}\ and\ \bibinfo {author} {\bibfnamefont {J.}~\bibnamefont
  {Nagler}},\ }\href@noop {} {\bibfield  {journal} {\bibinfo  {journal} {Nat.
  Phys.}\ }\textbf {\bibinfo {volume} {11}},\ \bibinfo {pages} {531} (\bibinfo
  {year} {2015})}\BibitemShut {NoStop}%
\bibitem [{\citenamefont {Sole}\ and\ \citenamefont
  {Montoya}(2001)}]{sole01_ecological}%
  \BibitemOpen
  \bibfield  {author} {\bibinfo {author} {\bibfnamefont {R.~V.}\ \bibnamefont
  {Sole}}\ and\ \bibinfo {author} {\bibfnamefont {M.}~\bibnamefont {Montoya}},\
  }\href@noop {} {\bibfield  {journal} {\bibinfo  {journal} {Proc. Roy. Soc.
  London Ser. B}\ }\textbf {\bibinfo {volume} {268}},\ \bibinfo {pages} {2039}
  (\bibinfo {year} {2001})}\BibitemShut {NoStop}%
\bibitem [{\citenamefont {Milo}\ \emph {et~al.}(2002)\citenamefont {Milo},
  \citenamefont {Shen-Orr}, \citenamefont {Itzkovitz}, \citenamefont {Kashtan},
  \citenamefont {Chklovskii},\ and\ \citenamefont {Alon}}]{milo2002network}%
  \BibitemOpen
  \bibfield  {author} {\bibinfo {author} {\bibfnamefont {R.}~\bibnamefont
  {Milo}}, \bibinfo {author} {\bibfnamefont {S.}~\bibnamefont {Shen-Orr}},
  \bibinfo {author} {\bibfnamefont {S.}~\bibnamefont {Itzkovitz}}, \bibinfo
  {author} {\bibfnamefont {N.}~\bibnamefont {Kashtan}}, \bibinfo {author}
  {\bibfnamefont {D.}~\bibnamefont {Chklovskii}}, \ and\ \bibinfo {author}
  {\bibfnamefont {U.}~\bibnamefont {Alon}},\ }\href@noop {} {\bibfield
  {journal} {\bibinfo  {journal} {Science}\ }\textbf {\bibinfo {volume}
  {298}},\ \bibinfo {pages} {824} (\bibinfo {year} {2002})}\BibitemShut
  {NoStop}%
\bibitem [{\citenamefont {Gastner}\ and\ \citenamefont
  {Newman}(2006)}]{gastner06_distribution}%
  \BibitemOpen
  \bibfield  {author} {\bibinfo {author} {\bibfnamefont {M.~T.}\ \bibnamefont
  {Gastner}}\ and\ \bibinfo {author} {\bibfnamefont {M.~E.~J.}\ \bibnamefont
  {Newman}},\ }\href@noop {} {\bibfield  {journal} {\bibinfo  {journal} {Phys.
  Rev. E}\ }\textbf {\bibinfo {volume} {74}},\ \bibinfo {pages} {016117}
  (\bibinfo {year} {2006})}\BibitemShut {NoStop}%
\bibitem [{\citenamefont {Memmesheimer}\ and\ \citenamefont
  {Timme}(2006)}]{memmesheimer06_designingNeural}%
  \BibitemOpen
  \bibfield  {author} {\bibinfo {author} {\bibfnamefont {R.-M.}\ \bibnamefont
  {Memmesheimer}}\ and\ \bibinfo {author} {\bibfnamefont {M.}~\bibnamefont
  {Timme}},\ }\href@noop {} {\bibfield  {journal} {\bibinfo  {journal} {Phys.
  Rev. Lett.}\ }\textbf {\bibinfo {volume} {97}},\ \bibinfo {pages} {188101}
  (\bibinfo {year} {2006})}\BibitemShut {NoStop}%
\bibitem [{\citenamefont {Ronellenfitsch}\ and\ \citenamefont
  {Katifori}(2016)}]{katifori16_optimal}%
  \BibitemOpen
  \bibfield  {author} {\bibinfo {author} {\bibfnamefont {H.}~\bibnamefont
  {Ronellenfitsch}}\ and\ \bibinfo {author} {\bibfnamefont {E.}~\bibnamefont
  {Katifori}},\ }\href@noop {} {\bibfield  {journal} {\bibinfo  {journal}
  {Phys. Rev. Lett.}\ }\textbf {\bibinfo {volume} {117}},\ \bibinfo {pages}
  {138301} (\bibinfo {year} {2016})}\BibitemShut {NoStop}%
\bibitem [{\citenamefont {Bala}\ and\ \citenamefont
  {Goyal}(2000)}]{bala00_noncooperative}%
  \BibitemOpen
  \bibfield  {author} {\bibinfo {author} {\bibfnamefont {V.}~\bibnamefont
  {Bala}}\ and\ \bibinfo {author} {\bibfnamefont {S.}~\bibnamefont {Goyal}},\
  }\href@noop {} {\bibfield  {journal} {\bibinfo  {journal} {Econometrica}\
  }\textbf {\bibinfo {volume} {68}},\ \bibinfo {pages} {1181} (\bibinfo {year}
  {2000})}\BibitemShut {NoStop}%
\bibitem [{\citenamefont {Jackson}\ and\ \citenamefont
  {Watts}(2002)}]{jackson02_network_evolution}%
  \BibitemOpen
  \bibfield  {author} {\bibinfo {author} {\bibfnamefont {M.}~\bibnamefont
  {Jackson}}\ and\ \bibinfo {author} {\bibfnamefont {A.}~\bibnamefont
  {Watts}},\ }\href@noop {} {\bibfield  {journal} {\bibinfo  {journal} {J.
  Econ. Theory}\ }\textbf {\bibinfo {volume} {106}},\ \bibinfo {pages} {265}
  (\bibinfo {year} {2002})}\BibitemShut {NoStop}%
\bibitem [{\citenamefont {Even-dar}\ and\ \citenamefont
  {Kearns}(2007)}]{even07_smallWorldGameTheory}%
  \BibitemOpen
  \bibfield  {author} {\bibinfo {author} {\bibfnamefont {E.}~\bibnamefont
  {Even-dar}}\ and\ \bibinfo {author} {\bibfnamefont {M.}~\bibnamefont
  {Kearns}},\ }in\ \href
  {http://papers.nips.cc/paper/3071-a-small-world-threshold-for-economic-network-formation.pdf}
  {\emph {\bibinfo {booktitle} {Advances in Neural Information Processing
  Systems 19}}},\ \bibinfo {editor} {edited by\ \bibinfo {editor}
  {\bibfnamefont {B.}~\bibnamefont {Sch\"{o}lkopf}}, \bibinfo {editor}
  {\bibfnamefont {J.~C.}\ \bibnamefont {Platt}}, \ and\ \bibinfo {editor}
  {\bibfnamefont {T.}~\bibnamefont {Hoffman}}}\ (\bibinfo  {publisher} {MIT
  Press, Cambridge, MA},\ \bibinfo {year} {2007})\ pp.\ \bibinfo {pages}
  {385--392}\BibitemShut {NoStop}%
\bibitem [{\citenamefont {Jackson}(2008)}]{jackson08_network_book}%
  \BibitemOpen
  \bibfield  {author} {\bibinfo {author} {\bibfnamefont {M.~O.}\ \bibnamefont
  {Jackson}},\ }\href@noop {} {\emph {\bibinfo {title} {Social and economic
  networks}}},\ Vol.~\bibinfo {volume} {3}\ (\bibinfo  {publisher} {Princeton
  University Press, Princeton, NJ},\ \bibinfo {year} {2008})\BibitemShut
  {NoStop}%
\bibitem [{\citenamefont {Atabati}\ and\ \citenamefont
  {Farzad}(2015)}]{atabati15_strategic}%
  \BibitemOpen
  \bibfield  {author} {\bibinfo {author} {\bibfnamefont {O.}~\bibnamefont
  {Atabati}}\ and\ \bibinfo {author} {\bibfnamefont {B.}~\bibnamefont
  {Farzad}},\ }\href@noop {} {\bibfield  {journal} {\bibinfo  {journal} {Comp.
  Soc. Netw.}\ }\textbf {\bibinfo {volume} {2}},\ \bibinfo {pages} {1}
  (\bibinfo {year} {2015})}\BibitemShut {NoStop}%
\bibitem [{\citenamefont {Schr\"oder}\ \emph {et~al.}(2018)\citenamefont
  {Schr\"oder}, \citenamefont {Nagler}, \citenamefont {Timme},\ and\
  \citenamefont {Witthaut}}]{schroeder18_individualDecisions}%
  \BibitemOpen
  \bibfield  {author} {\bibinfo {author} {\bibfnamefont {M.}~\bibnamefont
  {Schr\"oder}}, \bibinfo {author} {\bibfnamefont {J.}~\bibnamefont {Nagler}},
  \bibinfo {author} {\bibfnamefont {M.}~\bibnamefont {Timme}}, \ and\ \bibinfo
  {author} {\bibfnamefont {D.}~\bibnamefont {Witthaut}},\ }\href@noop {}
  {\bibfield  {journal} {\bibinfo  {journal} {Phys. Rev. Lett.}\ }\textbf
  {\bibinfo {volume} {120}},\ \bibinfo {pages} {248302} (\bibinfo {year}
  {2018})}\BibitemShut {NoStop}%
\bibitem [{\citenamefont {Nagler}\ \emph {et~al.}(2011)\citenamefont {Nagler},
  \citenamefont {Levina},\ and\ \citenamefont {Timme}}]{nagler11_single_links}%
  \BibitemOpen
  \bibfield  {author} {\bibinfo {author} {\bibfnamefont {J.}~\bibnamefont
  {Nagler}}, \bibinfo {author} {\bibfnamefont {A.}~\bibnamefont {Levina}}, \
  and\ \bibinfo {author} {\bibfnamefont {M.}~\bibnamefont {Timme}},\
  }\href@noop {} {\bibfield  {journal} {\bibinfo  {journal} {Nat. Phys.}\
  }\textbf {\bibinfo {volume} {7}},\ \bibinfo {pages} {265} (\bibinfo {year}
  {2011})}\BibitemShut {NoStop}%
\bibitem [{\citenamefont {Sabidussi}(1966)}]{sabidussi66_closeness}%
  \BibitemOpen
  \bibfield  {author} {\bibinfo {author} {\bibfnamefont {G.}~\bibnamefont
  {Sabidussi}},\ }\href {\doibase 10.1007/BF02289527} {\bibfield  {journal}
  {\bibinfo  {journal} {Psychometrika}\ }\textbf {\bibinfo {volume} {31}},\
  \bibinfo {pages} {581} (\bibinfo {year} {1966})}\BibitemShut {NoStop}%
\bibitem [{\citenamefont {Newman}(2010)}]{newman_2010_networks}%
  \BibitemOpen
  \bibfield  {author} {\bibinfo {author} {\bibfnamefont {M.~E.~J.}\
  \bibnamefont {Newman}},\ }\href@noop {} {\emph {\bibinfo {title} {Networks:
  An Introduction}}}\ (\bibinfo  {publisher} {Oxford University Press,
  Oxford},\ \bibinfo {year} {2010})\BibitemShut {NoStop}%
\bibitem [{\citenamefont {Freeman}(1977)}]{freeman77_betweeness}%
  \BibitemOpen
  \bibfield  {author} {\bibinfo {author} {\bibfnamefont {L.~C.}\ \bibnamefont
  {Freeman}},\ }\href@noop {} {\bibfield  {journal} {\bibinfo  {journal}
  {Sociometry}\ }\textbf {\bibinfo {volume} {40}},\ \bibinfo {pages} {35}
  (\bibinfo {year} {1977})}\BibitemShut {NoStop}%
\bibitem [{\citenamefont {van Dolder}\ and\ \citenamefont
  {Buskens}(2014)}]{vanDolder14_individual}%
  \BibitemOpen
  \bibfield  {author} {\bibinfo {author} {\bibfnamefont {D.}~\bibnamefont {van
  Dolder}}\ and\ \bibinfo {author} {\bibfnamefont {V.}~\bibnamefont
  {Buskens}},\ }\href@noop {} {\bibfield  {journal} {\bibinfo  {journal} {PloS
  one}\ }\textbf {\bibinfo {volume} {9}},\ \bibinfo {pages} {e92276} (\bibinfo
  {year} {2014})}\BibitemShut {NoStop}%
\bibitem [{\citenamefont {Bertsekas}(1998)}]{bertsekas98_optimization}%
  \BibitemOpen
  \bibfield  {author} {\bibinfo {author} {\bibfnamefont {D.~P.}\ \bibnamefont
  {Bertsekas}},\ }\href@noop {} {\emph {\bibinfo {title} {Network optimization:
  continuous and discrete models}}}\ (\bibinfo  {publisher} {Athena Scientific,
  Belmont},\ \bibinfo {year} {1998})\BibitemShut {NoStop}%
\bibitem [{\citenamefont {Watts}(2001)}]{watts01_dynamic}%
  \BibitemOpen
  \bibfield  {author} {\bibinfo {author} {\bibfnamefont {A.}~\bibnamefont
  {Watts}},\ }\href@noop {} {\bibfield  {journal} {\bibinfo  {journal} {Games
  Econ. Behav.}\ }\textbf {\bibinfo {volume} {34}},\ \bibinfo {pages} {331}
  (\bibinfo {year} {2001})}\BibitemShut {NoStop}%
\end{thebibliography}%

\end{document}